\documentclass[aps,pre,twocolumn,superscriptaddress,amsmath]{revtex4}
\usepackage{graphicx,amssymb,color}
\begin{document}

%\preprint{}

\title{Bubble relaxation dynamics in homopolymer DNA sequences}

\author{M.~Hillebrand} \email{m.hillebrand@uct.ac.za} \affiliation{Nonlinear Dynamics and Chaos Group, Department of Mathematics and
Applied Mathematics, University of Cape Town, Rondebosch 7701,
South Africa}
\author{G.~Kalosakas} \email{georgek@upatras.gr} \affiliation{Department of Materials Science, University of Patras, GR-26504 Rio, Greece}
\author{A.R.~Bishop}  \affiliation{Los Alamos National Laboratory,
Los Alamos, NM, 87545, United States}
\author{Ch.~Skokos} 
%\homepage{http://math\_research.uct.ac.za/~hskokos/}
\affiliation{Nonlinear Dynamics and Chaos Group, Department of Mathematics and
Applied Mathematics, University of Cape Town, Rondebosch 7701,
South Africa}

\date{\today}

%=========================
\begin{abstract}
Understanding the inherent timescales of large bubbles in DNA is critical to a thorough comprehension of its physicochemical characteristics, as well as their potential role on helix opening and biological function.
In this work we employ the coarse-grained Peyrard-Bishop-Dauxois model of DNA to study relaxation dynamics of large bubbles in homopolymer DNA, using simulations up to the microsecond time scale.
By studying energy autocorrelation functions of relatively large bubbles inserted into thermalised DNA molecules, we extract characteristic relaxation times from the equilibration process for both adenine-thymine (AT) and guanine-cytosine (GC) homopolymers.
Bubbles of different amplitudes and widths are investigated through extensive statistics and appropriate fittings of their relaxation. Characteristic relaxation times increase with bubble height and width.
We show that, within the model, relaxation times are two orders of magnitude longer in GC sequences than in AT sequences.
Overall, our results confirm that large bubbles leave a lasting impact on the molecule's dynamics, for times between $0.4-200$ns depending on the homopolymer type and bubble shape, thus clearly affecting long-time evolutions of the molecule.
\end{abstract}
%=========================

% \pacs{05.45.-a, ????}
% \keywords{}

\maketitle

%=========================
\section{Introduction} % (fold)
\label{sec:introduction}
The dynamics of biomolecules such as DNA have long been a source of interest, providing meaningful information beyond that yielded by the static molecular structure \cite{beece1980,ansari1985,sobel1985}.
In particular, the notion of extracting timescales for dynamical processes in DNA has attracted attention both theoretically and experimentally \cite{Somoza2004,Perez2007,Banerjee2007,Galindo2014,Parmar2016}, due to the importance of quantifying the impact of thermal and mechanical effects on the overall behaviour of the molecule.

A particularly interesting feature of DNA dynamics, which has been suggested to have a potential role in transcription and other biological processes, is the existence of local large base pair openings, often called bubbles, where regions of the double helix open.
These openings can be thermally-induced fluctuations, or promoted by base pair mismatching~\cite{Zeng2006}.
DNA bubbles can be experimentally studied through NMR experiments~\cite{leroy1988} and fluorescence spectra~\cite{Jose2012,Phelps2013}, as well as computationally using extensive molecular dynamics (MD) studies~\cite{Lavery2010,Galindo2014}.
It is also possible to study these breather-like excitations using nonlinear modelling (see e.g.~\cite{Peyrard2009} and references therein).

There have been many models of DNA proposed and studied in various detail~\cite{Manghi2016}, ranging from detailed \textit{ab initio} models~\cite{kaxiras12} to thermodynamically-motivated models~\cite{weber06} and empirical potentials accounting for the helical or curved structure of DNA~\cite{zoli13,depablo14,zoli18,zoli18b}, as well as free-energy-based methods~\cite{sun19}.
In this work we consider the Peyrard-Bishop-Dauxois (PBD) model~\cite{Peyrard1989,PBD,DPB93,Dauxois1995}, which provides an effective mesoscale view of DNA dynamics, successfully reproducing sharp denaturation curves and several experimental results~\cite{Campa1998}.
This model reduces the complex molecular structure of the double helix to a more tractable one dimensional lattice system, enabling the investigation of such diverse phenomena as intrinsic localised modes~\cite{Peyrard2000},  electronic transport where bubbles can cause charge trapping~\cite{Kalosakas2005,Kalosakas2011,Chetverikov2019}, chaoticity~\cite{Barre2001,Hillebrand2019}, DNA/TNA couplings~\cite{Muniz2020} and optical switching~\cite{Behnia2020}.
The incorporation of a sequence-dependent stacking parameter within the PBD model provides better accuracy with detailed denaturation (i.e.~complete separation of the double strand) results for a variety of DNA sequences exhibiting unusual melting behavior~\cite{Alexandrov2009ePBD}.

Using the PBD model, various studies of DNA breathing and fluctuational opening probabilities~\cite{Voulgarakis2004,Ares2005,Ares2007,Kalosakas2009,falo10} have been carried out.
Stretched exponential evolution has been found for the decay of equilibrium fluctuations of base pairs in DNA molecules \cite{Kalosakas2006}, exhibiting relaxation times beyond the picosecond scale.
Opening probability profiles and lifetimes of bubbles have also been studied extensively in the context of DNA promoters~\cite{Choi2004,Kalosakas2004,Alexandrov2006,Choi2008,Alexandrov2009,Alexandrov2010,Apostolaki2011,faloPRE12,huangJBE,faloPLOS,Hillebrand2021},
as well as more general bubble distributions for arbitrary sequences~\cite{Hillebrand2020}. 
These findings further indicated that more large bubbles can be distinguished in transcriptionally significant regions of DNA promoters than expected from average results, providing additional impetus to the interest of studying longer-time effects of openings in DNA molecules, in which times biological processes might be initiated.

Here, we further investigate bubble lifetimes' properties by studying relaxation times of large openings in the DNA double strand, such as may be produced by rare thermal fluctuations, induced by artificially engineered means, or naturally created by proteins.
Not only does this further the characterisation of DNA's response to large bubbles, but it provides a quantification of the time scales for the system's memory of out-of-equilibrium perturbations, and demonstrates that these large openings leave a long-lasting footprint on the dynamics of the molecule.

The paper is organised as follows.
In Section~\ref{sec:model_and_setup} we introduce the dynamical PBD model used here, along with the numerical methods, parameters, and the simulation protocols for investigating bubbles.
The results and analysis of data follow in Section~\ref{sec:results}, with the summary and conclusions in Section~\ref{sec:conclusions} closing out the report.

\section{Model and Setup} % (fold)
\label{sec:model_and_setup}

We perform dynamical simulations using the PBD model of DNA~\cite{PBD}, describing the molecule as a sequence of nonlinearly coupled base pairs.
The Hamiltonian function of the PBD model for a DNA sequence of $N$ base pairs, considering periodic boundary conditions is given by
\begin{equation}
    \label{eq:PBDHamiltonian}
    H = \sum_{n=1}^N  \left[ \frac{p_n^2}{2m}+V_1(y_n) + V_2(y_n,y_{n-1}) \right],
\end{equation}
with $y_0=y_N$.
The on-site energy interaction is governed by the Morse potential
\begin{equation}
    \label{eq:Morse}
    V_1(y_n) = D_n\left(e^{-a_n y_n} - 1\right)^2,
\end{equation}
and the nearest-neighbour stacking interaction is modelled by
\begin{equation}
    \label{eq:PBDStack}
    V_2(y_{n},y_{n-1}) = \frac{K_{n,n-1}}{2}\left(1+\rho e ^{b(y_n + y_{n-1})}\right)\left(y_n-y_{n-1}\right)^2.
\end{equation}
Here the $y_n$ are the displacements from equilibrium of each base pair, $p_n$ the corresponding momentum, $a_n$ and $D_n$ are the constants of the Morse on-site potential distinguishing AT or GC base pairs, while the coupling constants $K_{n,n-1}$ are sequence-dependent stacking strengths.
The parameter values used are $D_{GC}=0.075$eV, $a_{GC}=6.9$\AA$^{-1}$ for GC base pairs and $D_{AT} = 0.05$eV, $a_{AT} = 4.2$\AA$^{-1}$ for AT base pairs, $\rho=2$, $b=0.35$\AA$^{-1}$~\cite{Campa1998},  while for the $K_{n,n-1}$ values see Ref.~\cite{Alexandrov2009ePBD} or Table I of Ref.~\cite{Hillebrand2020}.
In this work we consider only the two homopolymer cases, so we have $K_{AA}=K_{TT}=0.0228$eV/\AA$^2$ for pure AT sequences, i.e.~poly(dA)$\cdot$poly(dT)
and $K_{GG}=K_{CC}= 0.0192$eV/\AA$^2$ for the GC case, poly(dG)$\cdot$poly(dC).
The temperature is kept at physiological level, around 310K, meaning different energies are used in the AT and GC cases due to our microcanonical constant-energy simulations.
Based on the energy-temperature relation in the considered PBD model~\cite{Hillebrand2020}, we use an average energy per particle of $\varepsilon_i=0.043$eV for AT sequences, while for GC sequences $\varepsilon_i=0.045$eV is used.

Importantly, we make use of symplectic integration techniques for the conservative Hamiltonian system~\cite{Hairer2002,DMMS19}, and specifically the symplectic Runge-Kutta-Nystr{\o}m method SRKNb6~\cite{Blanes2002}.
Not only is this method efficient and accurate, but it also preserves the system's symplectic nature, which  means that performing simulations even up to several microseconds is possible without sacrificing efficiency for precision, as would be required with typical non-symplectic schemes with growing errors.
A relative energy error of around $|H(t)-H(0)|/H(0)<10^{-8}$ is maintained throughout all simulations.

Each simulation is run starting from equilibrium displacements $y_i=0$, with a set of initial momenta drawn from a zero-mean random normal distribution, which are scaled to provide the correct total energy.
100 base pair long sequences are used, with periodic boundary conditions to minimise finite size effects.
For every case presented, 1000 simulations are used to ensure statistical robustness for the results.
The first 10ns of the evolution are taken as a thermalisation period, whereafter we directly introduce a bubble by replacing the central base pairs' displacements with a Gaussian-shaped initial perturbation.
This Gaussian has the form
\begin{equation}
    \label{eq:gaussian}
    y(x) = h \textrm{e}^{-\frac{(x-c)^2}{2\sigma^2}},
\end{equation}
where $h$ is the height of the bubble, $c$ is the centre located in the middle of the DNA sequence  (i.e. $c=50$ for our 100-base-pair sequences), and $\sigma$ the ``standard deviation'', characterising the width of the Gaussian.
The total number of base pairs which have their displacements replaced by this Gaussian is denoted $w$, which determines the width of the introduced bubble.
The parameter $\sigma$ in Eq.~(\ref{eq:gaussian}) is given by $\sigma = w/6$ to ensure that the inserted bubble has tails at near equilibrium displacement.
Therefore the out-of-equilibrium bubbles are characterised by their height $h$ and width $w$.
For all bubbles studied in this work, we keep $w$ as an odd number of base pairs to allow for even length tails on either side of the centre.
The displacements of the remaining base pairs (those not belonging to the bubble region) are rescaled using a bisection algorithm to retain the total original energy $H$ of the chain to within an accuracy of $|H^\prime - H|<10^{-10}$eV, where $H^\prime$ denotes the energy of the system after the bubble is inserted and displacements rescaled.
With the momenta remaining unaffected, the temperature is unchanged by the bubble insertion.

\begin{figure}[tb]
    \centering
    \includegraphics[width=0.9\columnwidth]{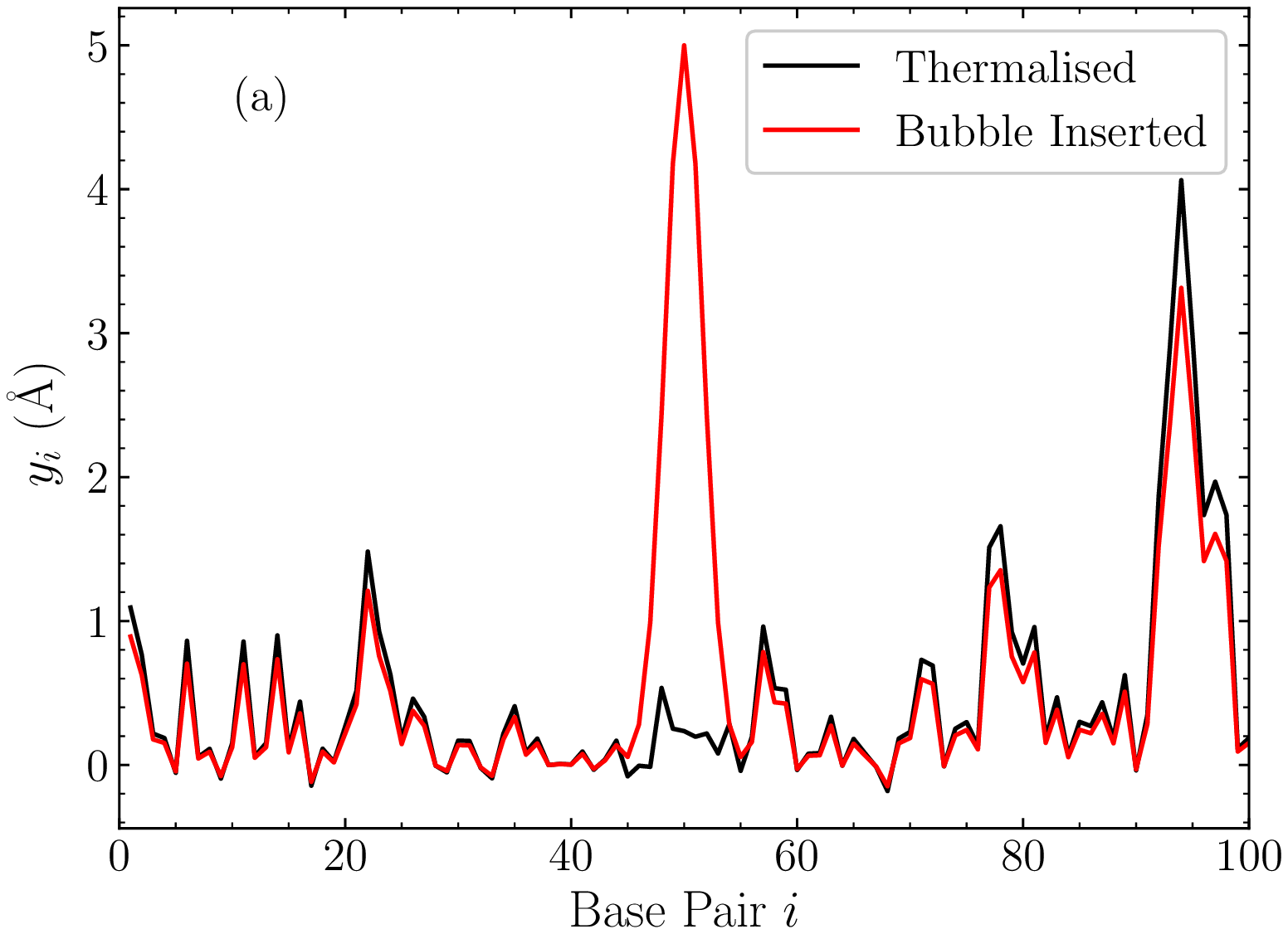}
    \includegraphics[width=0.9\columnwidth]{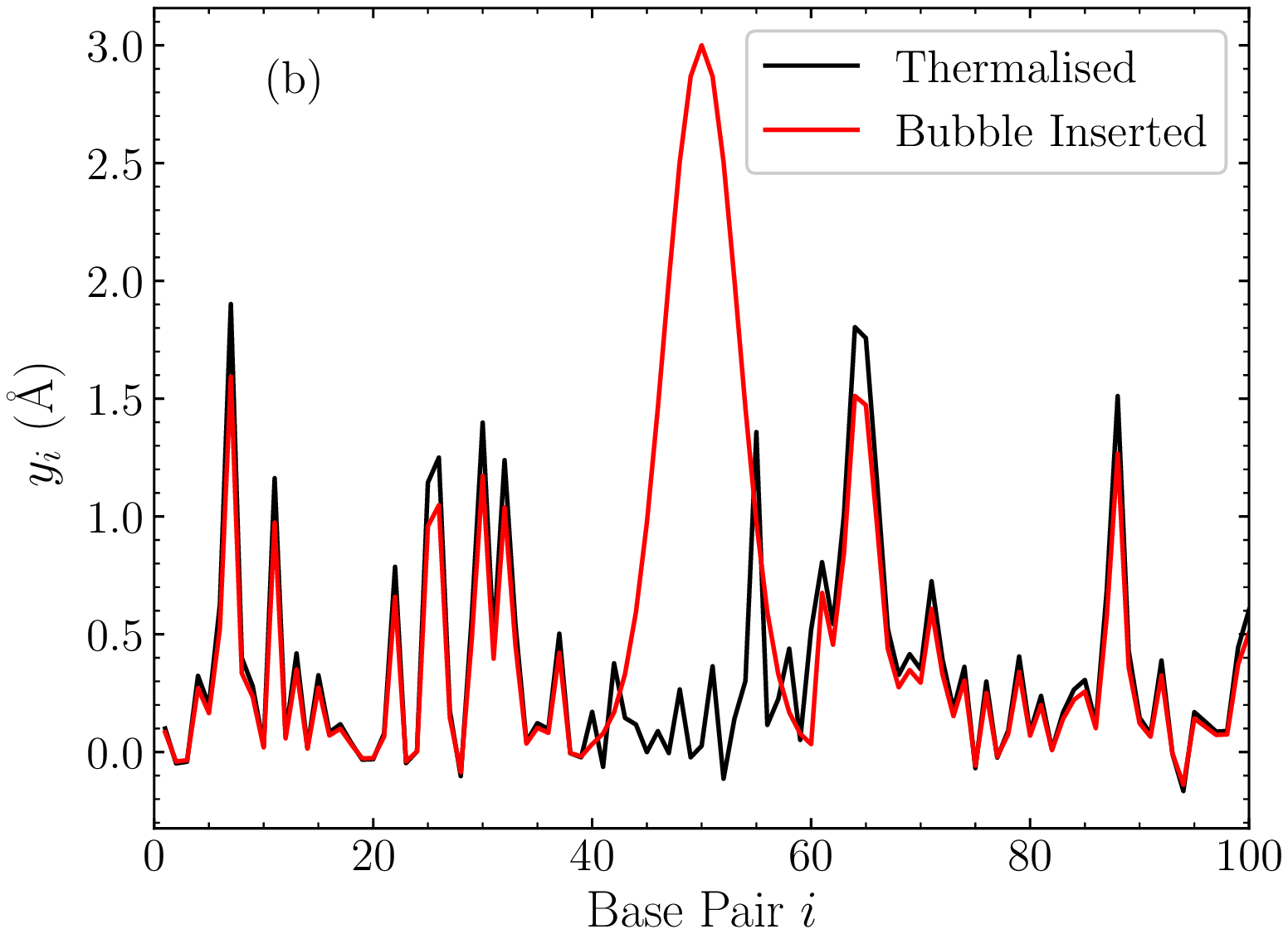}
    \caption{Two representative cases of the displacements in a thermalised homopolymer AT sequence at $t=10$ns (black), and the corresponding profile after the Gaussian perturbation is introduced and the remaining displacements rescaled (red). (a) Bubble with a width $w=11$ base pairs and height $h=5$\AA. (b) Bubble with $w=21$ base pairs and height $h=3$\AA.}
    \label{fig:initialInsertion}
\end{figure}
%%%

In Fig.~\ref{fig:initialInsertion} we illustrate the profiles of displacements within the DNA chain after the introduction of the bubble, along with the pre-insertion thermalised equilibrium, for two representative cases with bubble width $w=11$ base pairs and height 5\AA~[Fig.~\ref{fig:initialInsertion}(a)], and $w=21$ base pairs and height 3\AA~[Fig.~\ref{fig:initialInsertion}(b)] in a pure AT sequence.
These cases help to put the height and width of these bubbles into context,
confirming that these inserted Gaussians are generally out-of-equilibrium large perturbations.

To examine the relaxation of these non-equilibrium bubbles, we track the displacements as the molecule evolves through time, and compute autocorrelation functions for the bubble region.
The non-normalised displacement autocorrelation function is calculated as
\begin{equation}
    \label{eq:cd}
    C_D(t) =  \frac{1}{w}   \Biggl\langle \sum_{i=(N-w+1)/2}^{i=(N+w-1)/2}y_{\textit{i}}(0)y_{\textit{i}}(t) \,\Biggr \rangle,
\end{equation}
so the sum over $i$ spans the displacements within the region of interest, which is the $w$-base-pair region where the bubble is inserted.
The notation $\langle \cdot \rangle$ in Eq.~(\ref{eq:cd}) means that the final autocorrelation function is calculated averaging over the ensemble of 1000 simulations.
The limiting or asymptotic value of this correlation function is 
\begin{equation}
    \label{eq:cdLim}
    \chi_{D} = \left( \frac{1}{w}\sum_{i=(N-w+1)/2}^{i=(N+w-1)/2}y_\textit{i}(0) \right) y_\textit{eq},
\end{equation}
where $y_{eq}$ is the average thermal equilibrium displacement of the entire homopolymer sequence.
For this value of $y_{eq}$, we use an average of the displacements of all base pairs, from all simulations, after thermalisation and before the bubble is inserted.

In addition to the relaxation of the displacements, we can study the relaxation of the energy distribution induced by these bubbles.
To this end, we compute the energy at each base pair $i$ as
\begin{equation}
    \label{eq:ePerBP}
    \varepsilon_i = \frac{p_i^2}{2m}+V_1(y_i) + \frac{1}{2}\left[V_2(y_{i+1},y_i)+V_2(y_i,y_{i-1})\right],
\end{equation}
and study the autocorrelation function relaxation for this quantity.
For these energies per base pair $\varepsilon_i$ the autocorrelation function is given by
\begin{equation}
    \label{eq:ce}
    C_E(t) = \frac{1}{w} \left\langle \sum_{i=(N-w+1)/2}^{i=(N+w-1)/2}\varepsilon_{\textit{i}}(0)\varepsilon_{\textit{i}}(t) \right\rangle,
\end{equation}
where again the sum is over the base pairs inside the bubble region.
In the same fashion as for the displacements, the limiting value for the energy autocorrelation function is given as
\begin{equation}
     \label{eq:ceLim}
    \chi_{E} = \left( \frac{1}{w}\sum_{i=(N-w+1)/2}^{i=(N+w-1)/2}\varepsilon_\textit{i}(0) \right) \varepsilon_\textit{eq},
\end{equation} 
where in fact we have explicitly that the equilibrium energy per base pair is simply the initial energy per base pair. Thus, for GC sequences $\varepsilon_{eq}^{GC}=0.045$eV and for AT $\varepsilon_{eq}^{AT}=0.043$eV.

In order to compute the aforementioned autocorrelation functions in our numerical simulations, the values of the displacements and energies per base pair in the bubble region are recorded starting from the insertion time, which is considered to be $t=0$.
The simulations are run for a further 1$\mu$s for AT sequences, and 5$\mu$s for GC sequences, and data stored in log time.
These recording times were found to be sufficient for the system to exhibit a complete relaxation of the autocorrelation functions towards equilibrium.
%%%%%%%%%%%%%%%%%%%%%%

\section{Results and discussion} % (fold)
\label{sec:results}
\begin{figure*}[tb]
    \centering
    \includegraphics[width=\textwidth]{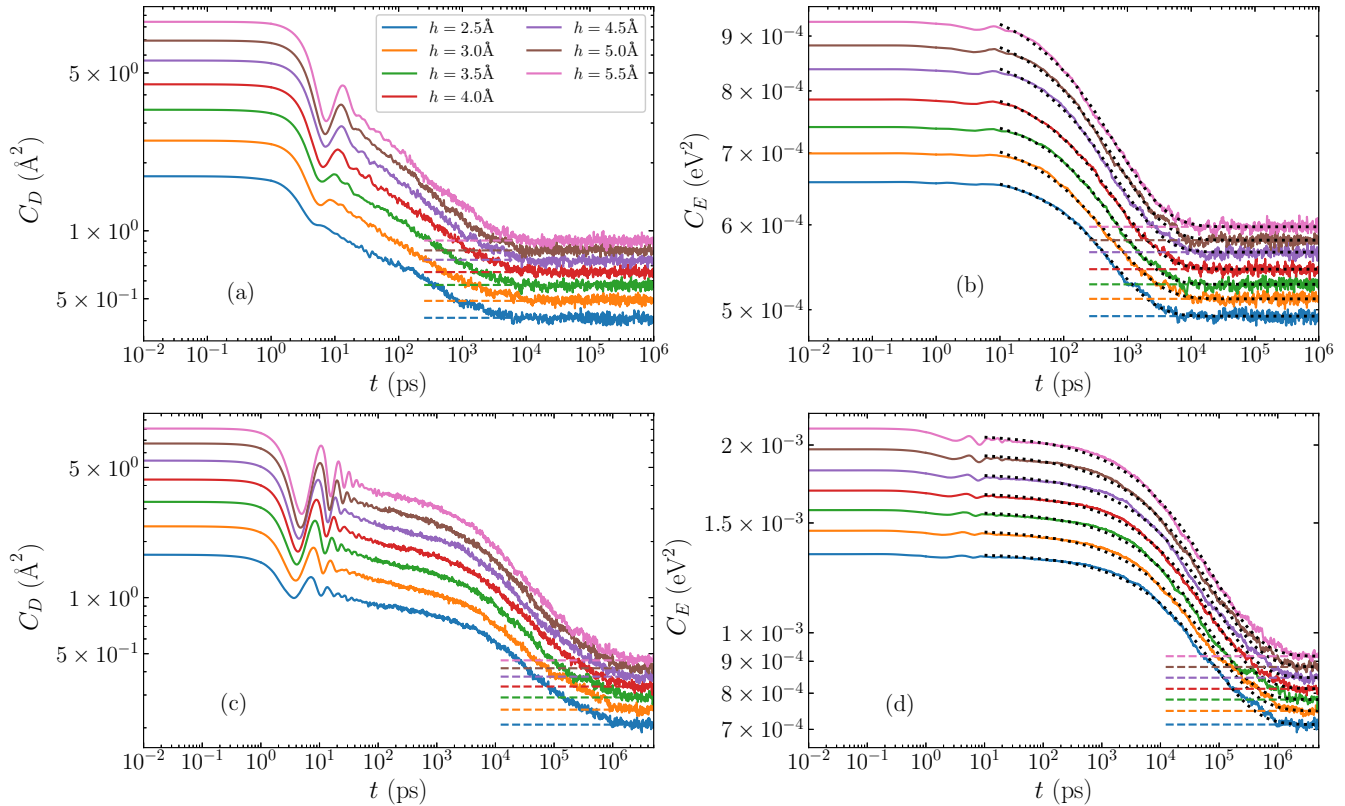}
    \caption{Time evolution of the average autocorrelation functions for homogeneous AT and GC sequences and different initial bubble heights ranging from $h=2.5$\AA\ to $h=5.5$\AA. Top row: AT sequences, for a bubble width of $w=19$ base pairs; (a) displacement autocorrelation functions $C_D(t)$, [Eq.~\eqref{eq:cd}], and (b) energy autocorrelation functions $C_E(t)$, [Eq.~\eqref{eq:ce}].
    Bottom row: GC sequences with a bubble width of $w=11$ base pairs; (c) displacement and (d) energy autocorrelation functions.
    In all cases the dashed lines mark the expected limiting values $\chi_D$ and $\chi_E$ from Eqs.~\eqref{eq:cdLim} and~\eqref{eq:ceLim}, respectively. Dotted lines in (b) and (d) represent stretched exponential fittings with Eq.~\eqref{eq:stretch}. 
    }
    \label{fig:dataplots}
\end{figure*}

Illustrative cases of individual autocorrelation functions are depicted in log-log scale in Fig.~\ref{fig:dataplots}.
For the poly(dA)$\cdot$poly(dT) sequence [Figs.~\ref{fig:dataplots}(a) and(b)], we show the relaxation of a bubble of width $w=19$ base pairs, and various heights ranging from $h=2.5$\AA\ to $h=5.5$\AA.
The displacement autocorrelation functions $C_D(t)$ [Eq.~\eqref{eq:cd}] are shown in Fig.~\ref{fig:dataplots}(a), and the energy autocorrelation functions $C_E(t)$ [Eq.~\eqref{eq:ce}] in Fig.~\ref{fig:dataplots}(b).
The poly(dG)$\cdot$poly(dC) case is demonstrated with a bubble of width $w=11$ base pairs, and the same range of heights $h=2.5-5.5$\AA.
Figure~\ref{fig:dataplots}(c) gives the displacement autocorrelation functions $C_D$, with the energy counterpart $C_E$ seen in Fig.~\ref{fig:dataplots}(d).
In all plots, for each height the expected limiting values, provided by Eqs.~\eqref{eq:cdLim} and \eqref{eq:ceLim}, are indicated  by horizontal dashed lines, to which the corresponding autocorrelation functions eventually converge.
The smoothness of the curves (even in log scale) confirms that we have sufficient statistics to eliminate large deviations in the data set.

A typical relaxation process of the AT homopolymers as demonstrated by the displacement autocorrelation function $C_D$ is seen in Fig.~\ref{fig:dataplots}(a).
Here we see two distinct stages of the relaxation of the autocorrelation function; an initial oscillatory region until around $t=20$ps, with increasing amplitude as the bubble height grows, followed by a steady rapid decay towards the equilibrium value, which is generally reached after several nanoseconds.
The energy autocorrelation function $C_E$ [Fig.~\ref{fig:dataplots}(b)] also exhibits two stages of decay, but the oscillations for the first picoseconds, coinciding with the window of oscillations in the $C_D$ functions, are significantly suppressed.
The $C_E$ autocorrelations reach the limiting value slightly later than the $C_D$ ones.

In the case of GC sequences [Figs.~\ref{fig:dataplots}(c) and (d)], similar but distinct decreasing behaviours of the autocorrelation functions are clearly visible.
First there is the initial oscillatory period up to times $t=10-40$ ps depending on height, where coherent peaks and troughs are especially visible in the displacement autocorrelation functions [Fig.~\ref{fig:dataplots}(c)], and less in the large-height energy functions [\ref{fig:dataplots}(d)].
In the GC homopolymers there are more oscillations during the first oscillatory stage of the relaxation in comparison to the AT sequences. When the oscillations diminish the slow decay of the
autocorrelation functions continues up to several nanoseconds in both $C_D$ and $C_E$, which then gives way to the second stage of the faster decay process lasting until around 1$\mu$s, when equilibrium values are reached.

\begin{figure*}
    \centering
    \includegraphics[width=0.49\textwidth]{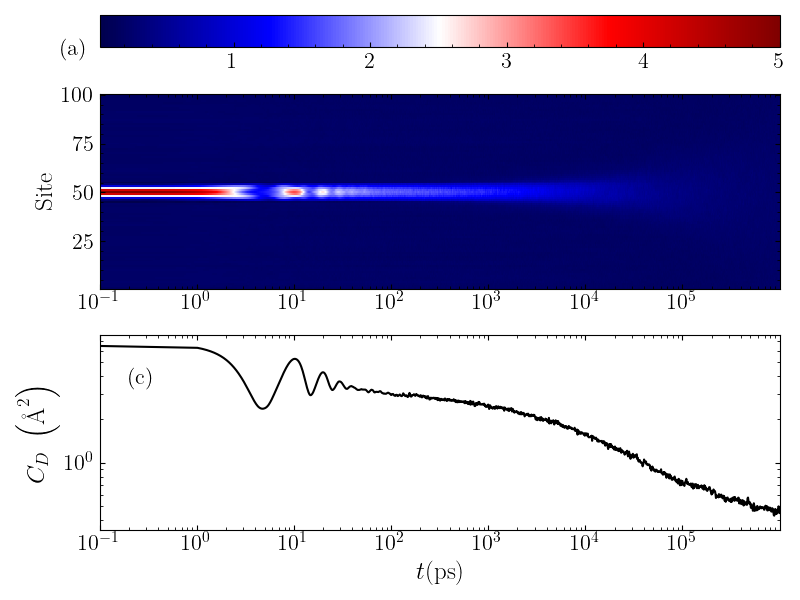}
    \includegraphics[width=0.49\textwidth]{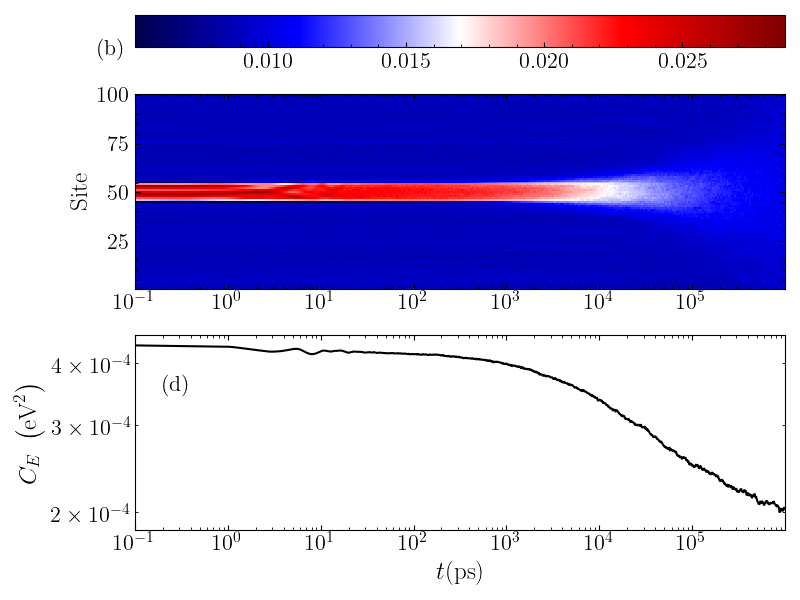}
    \includegraphics[width=0.4\textwidth]{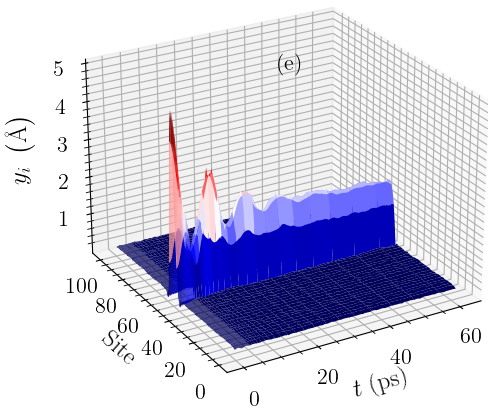}
    \hspace{2cm}
    \includegraphics[width=0.4\textwidth]{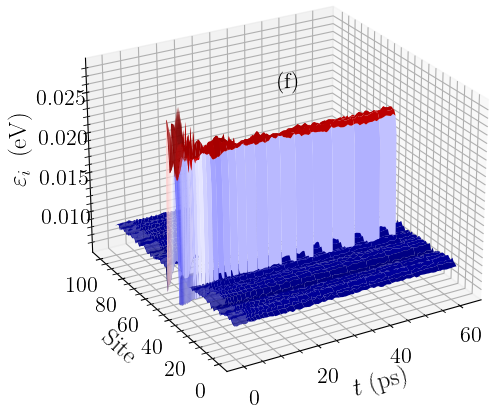}
    \caption{The evolution of the averaged displacement and energy density profiles through time, for the case of a GC homopolymer with an initial bubble of height $h=5$\AA, and width $w=11$ base pairs. The left column shows data for the displacements, and the right for the energy densities. Panels (a) and (b) show a density plot of the long-time evolution profile of the displacements and energy densities respectively, with the intensity labelled according to the colorbar above the panels. The corresponding average autocorrelation functions, $C_D$ and $C_E$, are shown in panels (c) and (d) respectively. A 3D representation of the early stage of the evolution is shown in panels (e) and (f) for the displacements and energy densities, respectively.}
    \label{fig:colourplots}
\end{figure*}

In order to find the mechanism responsible for these distinct relaxations, we can directly visualise the time evolution of the bubble displacements and energy densities averaged over one thousand realisations.
In Fig.~\ref{fig:colourplots}, typical evolution profiles are illustrated for a GC homopolymer with an initial bubble of height $h=5$\AA\ and width $w=11$ base pairs.
Considering first the displacement profile shown in a density plot in Fig.~\ref{fig:colourplots}(a), we see that the oscillations in the $C_D$ autocorrelation [Fig.~\ref{fig:colourplots}(c)] correspond exactly to large oscillations in the base pair displacements within the bubble region.
These oscillations appear as a result of the rearrangement of the initial perturbation towards a more stable localised structure, which at this width ($w=11$ base pairs) takes a peak height just below 2\AA, as can be seen in the 3D depiction of this stage of evolution in Fig.~\ref{fig:colourplots}(e).
The large oscillations are seen until around $t=30$ps, whereafter the displacement profile exhibits a nearly constant-height structure in the bubble region.
The resulting, more stable bubble then slowly decays giving rise to the ``slower'' relaxation of the autocorrelation function visible in Fig.~\ref{fig:colourplots}(c) and earlier observed in Fig.~\ref{fig:dataplots} until several nanoseconds. Then the second stage of relaxation follows, which is characterised by a gradual spreading and the complete disappearance of the bubble, leading to complete equilibration [see Fig.~\ref{fig:colourplots}(a)], a process signified by the rapid decay of the autocorrelation function in the nanosecond to microsecond time scale [Figs.~\ref{fig:dataplots} and~\ref{fig:colourplots}(c)].

The energy densities show a similar behaviour, but on a more muted scale as regards the initial oscillations during the bubble rearrangement at the first stage of relaxation.
As demonstrated in Fig.~\ref{fig:colourplots}(b), the initial oscillations are much smaller for the energy profile, and the rearranged localised structure remains almost stable for times up to nanoseconds.
This is accompanied by a correspondingly flatter autocorrelation function [Fig.~\ref{fig:colourplots}(d)] for this period, before the complete thermalisation towards equilibrium. The 3D visualisation in Fig.~\ref{fig:colourplots}(f) shows the early oscillations followed by a constant-height energy profile in the bubble region, with energy density per base pair just above 0.2eV. Also in this case, the rapid decay of $C_E$ at the second, final stage of the relaxations denotes the spreading and disappearance of the bubble [see Figs.~\ref{fig:colourplots}(b) and~(d)].

These observations are consistent to the behavior depicted in Fig.~\ref{fig:dataplots} that for a fixed width a greater height of the initial bubble perturbation results in larger initial oscillations of the autocorrelation functions.
Since these oscillations are produced by the rearrangement process mentioned above, the closer the initial bubble is to the nearly stable localised structure corresponding to its width, the smaller the changes required for the initial profile to be adapted to the inherent state.
These smaller rearrangements are reflected as weaker oscillations in the autocorrelation functions, especially in the $C_D$ functions where the oscillations are more evident.

It seems that both displacement and energy autocorrelation functions for AT sequences in Figs.~\ref{fig:dataplots}(a) and~\ref{fig:dataplots}(b) show the oscillatory region and the rapid decaying second stage, while the slow decaying relaxation behaviour is not evident, in contrast to what happens in the GC case.
While not shown here, we have found that in AT homopolymers the inherent rearranged bubble has a dramatically shorter lifetime, corresponding to the lack of the ``slow'' relaxation region in this case [see Figs.~\ref{fig:dataplots}(a) and~(b)].
Because of this, the time required for a complete equilibration, signifying the loss of any memory about the initial perturbation,
is around several nanoseconds for the AT homopolymer, which is orders of magnitude faster than the GC relaxation.

The decaying relaxation process depicted in these log-scale plots is very suggestive of a stretched exponential behaviour.
We thus consider a fit of the decaying stage of the autocorrelation functions with a stretched exponential of the form
\begin{equation}
    \label{eq:stretch}
    C(t) = A\exp\left(-\left(\frac{t}{\tau}\right)^\beta\right) + \chi,
\end{equation}
where $\chi$ corresponds to the limiting value of the relevant autocorrelation function, while $A$, $\beta$ and $\tau$ are free parameters denoting the preexponential coefficient, the characteristic time constant and the stretched exponent respectively.

The dotted black lines in Fig.~\ref{fig:dataplots}(b) and (d) illustrate the stretched exponential fitted to the decay of the $C_E$ data, where the fitting has started from the time where all data have entered the smooth decay relaxation stage (i.e.~beyond the initial oscillations), corresponding to an initial time of around $t=10$ps.
This results in visibly good fits, matching the data very well in all cases, and accurately capturing the decaying process.
Critically, the good agreement with the data means that the fitted stretched exponential reaches the equilibrium value simultaneously with the measured data.

We note that the decay of the $C_D$ data can also be fitted with stretched exponentials.
However, the $C_E$ curves provide a much more robust and consistent behaviour. Especially in the AT case the $C_D$ fitting is very sensitive to the starting point of the fitting due to the relatively small decaying region after the oscillatory relaxation.
Consequently, and since the time scales of both energy and displacement relaxations are similar (even visible in the autocorrelation functions themselves in Fig.~\ref{fig:dataplots}), in this work we focus on the relaxation timescales of the energy autocorrelation functions.

\begin{figure*}[tb]
    \centering
    \includegraphics[width=\columnwidth]{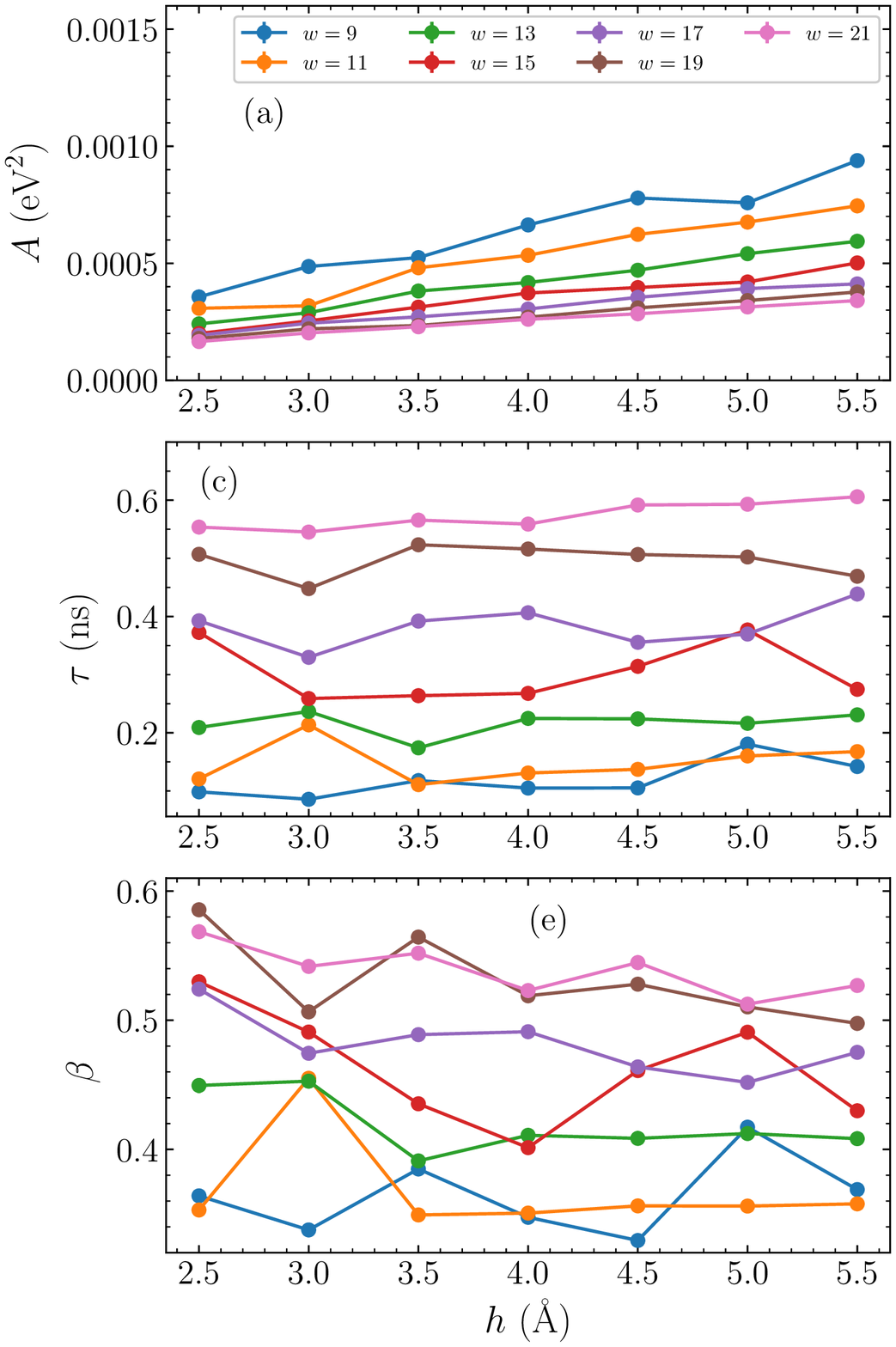}
    \includegraphics[width=\columnwidth]{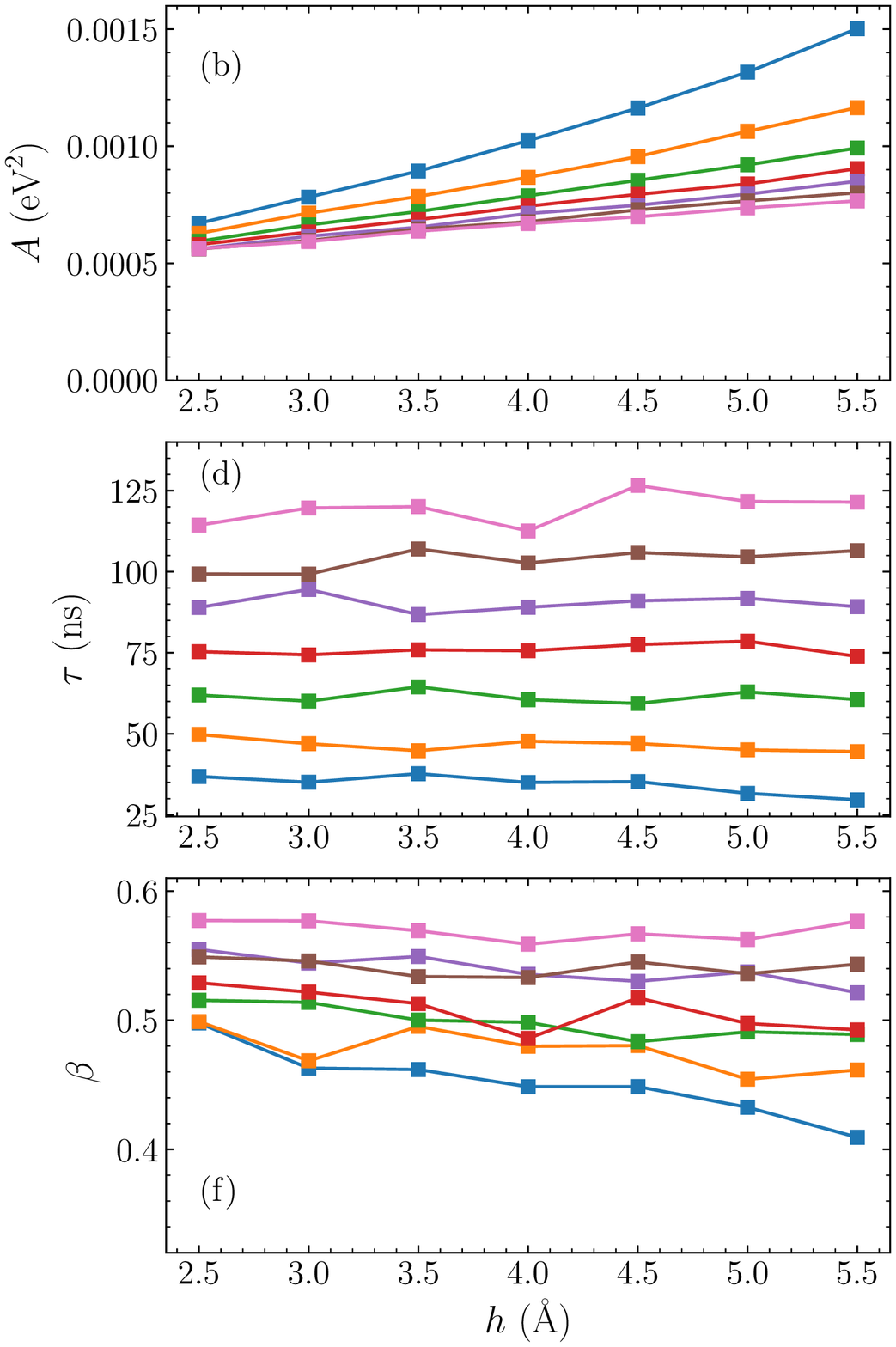}
    \caption{The stretched exponential parameters of Eq.~\eqref{eq:stretch}, as fitted to the energy autocorrelation functions $C_E(t)$ [Eq.~\eqref{eq:ce}], for varying heights $h$ and widths $w$ of the initial bubble perturbation. The left column shows the parameters for AT sequences, and the right column for GC sequences. (a) and (b): The preexponential coefficient $A$. (c) and (d): The characteristic time constant $\tau$. (e) and (f): The stretched exponent $\beta$.  The line connections are used to guide the eye.}
    \label{fig:params}
\end{figure*}

Having established this stretched exponential behaviour, we can now consider the dependence of the fitting parameters on the physical characteristics of the bubble perturbation -- its height $h$ and width $w$.
Performing this fit for a series of widths $w$ between 9 and 21 base pairs, and taking heights $h$ between 2.5 and 5.5\AA, we find the parameters of Eq.~\eqref{eq:stretch} in each case. The results are shown in Fig.~\ref{fig:params} as a function of $h$ with the different widths represented as separate data sets.
The parameters for AT homopolymers are shown in the left column, Figs.~\ref{fig:params}(a), (c) and (e), and for the GC sequences on the right, Figs.~\ref{fig:params}(b), (d) and (f).

Generally, the preexponential parameter $A$ increases consistently with height [Figs.~\ref{fig:params}(a) and (b)], corresponding to the visible height-dependence of the initial $C_E$ values, subtracting the corresponding $\chi_E$, seen in Figs.~\ref{fig:dataplots}(b) and (d).
When the width increases however, $A$ reduces, with the difference between the initial value of the autocorrelation function and its equilibrium value decreased.

Progressing through the parameters, we observe that the characteristic time $\tau$ [Figs.~\ref{fig:params}(c) and (d)] is hardly affected by the height of the initial bubble, in accordance to the results of Figs.~\ref{fig:dataplots}(b) and (d).
There are small, non-systematic variations with height, but particularly the GC parameters in Fig.~\ref{fig:params}(d) remain around the same value as height changes.
The dependence on width on the other hand is very clear, with a steady decrease in $\tau$ values as the bubble width shrinks.
This parameter most starkly reflects the difference in the relaxation dynamics between the AT and GC homopolymers.
While the $A$ [Figs.~\ref{fig:params}(a) and (b)] and $\beta$ [Figs.~\ref{fig:params}(e) and (f)] parameters are of comparable size between the two homopolymers, the $\tau$ values in GC sequences are more than two orders of magnitude greater than their AT counterparts.
This reflects a slower relaxation dynamics in the GC case.
We note that while the notion of a ``large displacement'' is not the same for AT and GC base pairs -- the energy required to stretch an AT base pair to a certain displacement is much lower than the energy required to stretch a GC base pair to the same displacement -- this discrepancy is certainly not on the scale of orders of magnitude.
%even comparing the largest AT displacements ($h=5.5$\AA) with the smallest GC displacements ($h=2.5$\AA), we see this order of magnitude difference being clearly evident.
%Thus, the difference in timescales is clearly inherent to the base pair type, not merely reflecting the different length scales from the Morse potential Eq.~\eqref{eq:Morse}.

The final parameter, the stretched exponent $\beta$ [Figs.~\ref{fig:params}(e) and (f)], actually gently decreases with height, implying that the shape of the relaxation does in fact change slightly with the bubble height. 
While there are a few slightly inconsistent points, especially in the AT case, the overall decreasing trend is nevertheless evident.

We note that in the autocorrelation function fittings, uncertainties are estimated using a bootstrapping approach on the $C_E$ data points. The resultant errorbars in Fig.~\ref{fig:params} are in fact smaller than the markers in the plots, suggesting that these fits are clearly optimal for each autocorrelation function.

The generally systematic behaviour of the fitting parameters suggests that there is a distinct trend in the behaviour of the overall relaxation dynamics as the height and width of the initial bubble change.
In order to understand this overall trend, we make use of the characteristic average time $\tau_{av}$ of the stretched exponential of Eq.~\eqref{eq:stretch}, which can be computed as \cite{Ngai1984,Leon1997,Kalosakas2006}
\begin{equation}
    \label{eq:tav}
    \tau_{av} = \frac{\Gamma\left(1/\beta\right)}{\beta}\tau,
\end{equation}
with $\Gamma(x)$ being the conventional gamma function.
This average time can be interpreted in this context as a measure of the relaxation time for the initial out-of-equilibrium perturbation, and provides an overall quantification of the time scale for the system's memory of a bubble with particular height $h$ and width $w$.
This quantity not only provides a relaxation time for the studied bubbles, but $\tau_{av}$ also enables us to use a single number as a descriptor for each autocorrelation function.
So in this sense, we reduce the complexity of many autocorrelation functions like those in Fig.~\ref{fig:dataplots} to one number for each curve, enabling a much more effective comparison between relaxations for different initial bubbles.

\begin{figure}[tb]
    \centering
    \includegraphics[width=\columnwidth]{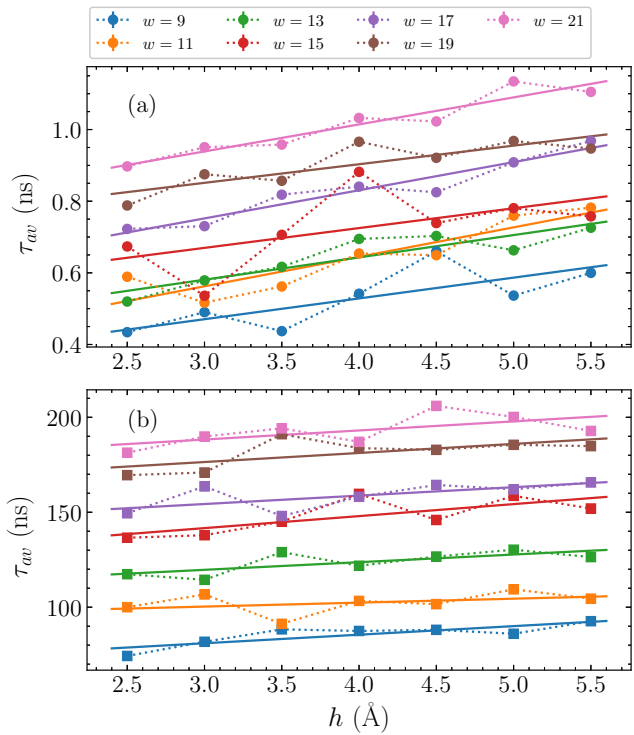}
    \caption{The average time $\tau_{av}$, Eq.~\eqref{eq:tav}, of the stretched exponential fitted to the autocorrelation functions $C_E(t)$ [Eq.~\eqref{eq:ce}], obtained from the parameters shown in Fig.~\ref{fig:params}, (a) for the AT sequences and (b) for the GC sequences (points). The dotted lines guide the eye for point-connection. The average times themselves are fitted by linear functions [see Eq.~\eqref{eq:tavFits}], depicted by  solid lines in (a) and (b).}
    \label{fig:tav}
\end{figure}

Based on the fitting parameters shown in Fig.~\ref{fig:params}, the average time $\tau_{av}$, computed though Eq.~\eqref{eq:tav}, is plotted in Fig.~\ref{fig:tav}, for each bubble width and height considered here.
The first panel, Fig.~\ref{fig:tav}(a), shows $\tau_{av}$ against the bubble height $h$ for the AT homopolymers, with each different width represented by a differently coloured series of dotted-line-connected points.
It is readily apparent that there is a consistent increase in the average relaxation time $\tau_{av}$ as the height $h$ grows, across all widths.
Concerning the bubble width, there is certainly a clear indication that the wider bubbles have longer timescales,
with the $w=21$ base pair bubbles [pink points in Fig.~\ref{fig:tav}(a)] having $\tau_{av}$ values larger than those seen for narrower $w=9$ base pair bubbles [blue points in Fig.~\ref{fig:tav}(a)], and a systematic change in times between.
The general relaxation timescales for the bubbles discussed here remain on the order of $10^2$ to $10^3$~ps.

Let us now consider the average times for the GC sequences, which are depicted  in Fig.~\ref{fig:tav}(b). Based on the observations of $\tau$ in Figs.~\ref{fig:params}(c) and (d), we expect a distinction between the relaxation time scales for the GC and AT sequences on the order of more than two orders of magnitude.
This difference is clearly seen in comparing Figs.~\ref{fig:tav}(a) and (b).
Despite this difference in magnitudes however, the same relative trends are visible that we saw in Fig.~\ref{fig:tav}(a).
More specifically, $\tau_{av}$ increases steadily with both height and width. The $w=21$ base pair bubbles relax in more than twice the time than the $w=9$ base pair bubbles.

With the characterisation of the relaxation dynamics through the average time $\tau_{av}$, and the systematic behaviour exhibited by these values in Fig.~\ref{fig:tav}, we are able to further quantify the effect of the bubble height and width on the relaxation time.
In particular, by fitting the data of Fig.~\ref{fig:tav} with a straight line of the form
\begin{equation}
    \label{eq:tavFits}
    \tau_{av} = \tau_0 + \alpha h,
\end{equation}
for each width, we can find the lines of best fit for both the AT and GC cases, and estimate a value for the intercept $\tau_0$ and the slope $\alpha$ depending on the width $w$.
These fits are shown by the solid lines in Fig.~\ref{fig:tav}, where in all cases the straight line provides a good description of the trend, thus accurately capturing the dependence of $\tau_{av}$ on the bubble height $h$.
Consequently, for a fixed width $w$, given the parameters $\tau_0$ and $\alpha$, it is possible to predict the typical relaxation time for a bubble of height $h$.

\begin{figure*}[tb]
    \centering
    \includegraphics[width=1.7\columnwidth]{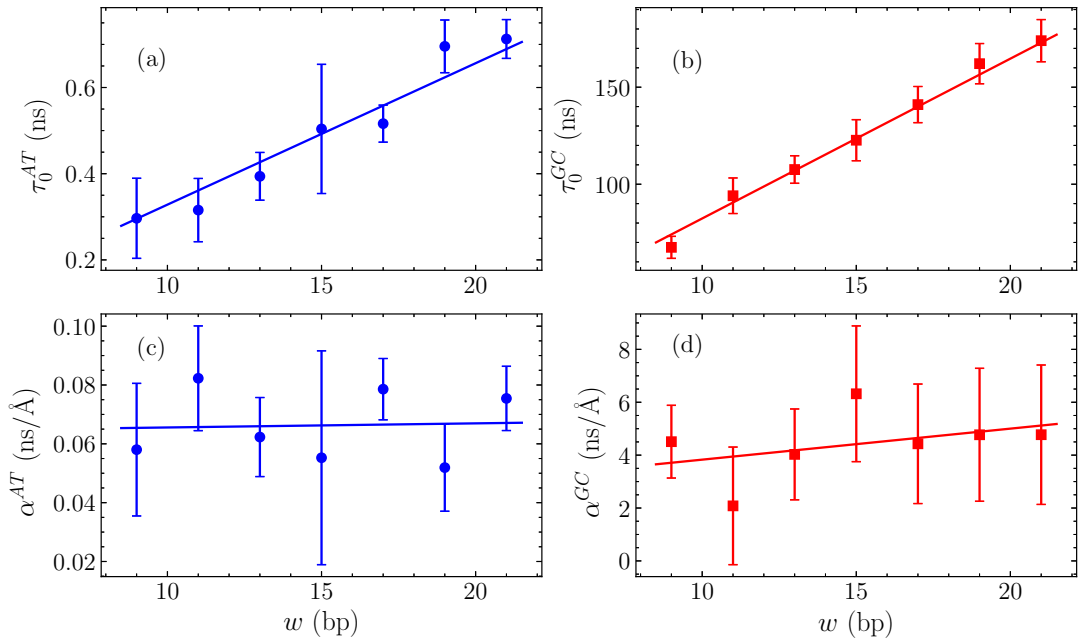}
    \caption{The variation of the parameters of linear fittings in $\tau_{av}$ [see the straight lines shown in Fig.~\ref{fig:tav} and Eq.~\eqref{eq:tavFits}] with the width $w$ measured in number of base pairs (bp). The vertical intercept $\tau_0$ is shown in (a) for AT sequences and in (b) for GC sequences, while the slope $\alpha$ is presented in (c) for AT homopolymers and in (d) for the GC ones (points). The data in (a) and (b) are in turn fitted with an origin-crossing straight line, while those in (c) and (d) by general linear fits (continuous lines). } 
    \label{fig:tavParams}
\end{figure*}

The calculated fitting  parameters $\tau_0$ and $\alpha$ are shown in Fig.~\ref{fig:tavParams} for the GC and AT homopolymers, as functions of the bubble width $w$.
In Fig.~\ref{fig:tavParams}(a) we see the intercepts for the AT case, $\tau_0^{AT}$, while Fig.~\ref{fig:tavParams}(b) displays the GC intercepts $\tau_0^{GC}$.
Both of these cases show a monotonic near-linear increase of $\tau_0$ with $w$, corresponding to the apparently steady increase with width in the average relaxation time $\tau_{av}$ visible in Fig.~\ref{fig:tav}.
The relatively smaller errorbars (here calculated through the covariance of the parameters) in the GC sequences convey the more stable, systematic behaviour of the fitting parameters in these homopolymers, in contrast to their AT counterparts (see Figs. \ref{fig:params} and \ref{fig:tav}).
The gradients $\alpha^{AT}$ and $\alpha^{GC}$ are given in Figs.~\ref{fig:tavParams}(c) and (d) for the AT and GC sequences respectively, still as functions of the bubble width $w$.
These values change significantly less with width than the intercepts, suggesting that the overall slope $\alpha$ is weakly dependent on $w$.

The variation of $\tau_0$ and $\alpha$ parameters depicted in Fig.~\ref{fig:tavParams} enables us to complete the quantification of bubble relaxation times as a function of both the height $h$ and width $w$ by fitting $\tau_0$ and $\alpha$ in turn with straight lines, shown as solid lines in all panels of Fig.~\ref{fig:tavParams}.
For the intercepts $\tau_0$ in Figs.~\ref{fig:tavParams}(a) and (b), a straight line through the origin was fitted finding $\tau_0^{GC}=8.2(0.1)w$ ns for GC homopolymers and $\tau_0^{AT}=0.032(0.001)w$ ns for AT. 
The numbers in parentheses denote the standard uncertainty in the last significant figure.
For both homopolymers, this fit provides an accurate quantification of the dependence on $w$.
The corresponding slopes $\alpha$ in Figs.~\ref{fig:tavParams}(c) and (d) are fitted by general linear fits, resulting in $\alpha^{GC}=0.1(0.1)w + 3(2)$ ns/\AA\  and $\alpha^{AT}=0.000(0.001)w + 0.06(0.02)$ ns/\AA\ for the GC and AT sequences, respectively, thus indicating a weak dependence on the bubble width.

Therefore, the characteristic relaxation times of AT and GC homopolymers exhibit generally linear dependence on both bubble height $h$ and width $w$, described approximately through the relations
\begin{equation}
    \label{eq:tavGC}
    \tau_{av}^{GC} = 8.2 w + (0.1 w + 3) h
\end{equation}
\begin{equation}
    \label{eq:tavAT}
    \tau_{av}^{AT} = 0.032 w + 0.06 h
\end{equation}
where in these equations the bubble heights are in \AA, the widths in base pairs, and the average relaxation times in ns.
The overall trend in the average relaxation times $\tau_{av}$ is the same for both AT and GC sequences, with the GC sequences relaxing around two orders of magnitudes more slowly than the AT sequences.

The calculated relaxation times of less than microseconds in all cases are consistent with previous detailed molecular dynamics simulations, finding that the short-time dynamics of DNA are only evident up to 1-5 $\mu$s~\cite{Galindo2014}. 
%Significantly, we also note that the average relaxation times for bubbles in the GC homopolymers [Fig.~\ref{fig:tav}(b)] are in the range of $80-200$ns, commensurate with the findings of Ref.~[\onlinecite{Galindo2014}] using molecular force field computations for short heteropolymers where average correlation functions of fluctuations were found to decay significantly in the time range of $80-130$ns.
%This agreement between results using few simulations with a high level of theory and our results using many simulations with a coarse-grained model is highly encouraging, and further motivates the use of such mesoscale approximations for exploring these long-time dynamics, which are difficult to probe with more detailed molecular dynamics.
However, note that the latter computations concern equilibrium fluctuations, rather than relaxation of out-of-equilibrium perturbations, as considered here.

In a previous work within the same PBD framework, using identical parameters as here apart from the fact that a common stacking parameter $K=0.025$ eV/\AA$^2$ is considered for both AT and GC sequences (which is close to the $K_{AA}=K_{TT}$ force constant used in this work, but more than  25\% larger than the corresponding $K_{GG}=K_{CC}$ value), characteristic rates for the decay of local displacement and energy autocorrelation functions of equilibrium fluctuations were computed at various temperatures~\cite{Kalosakas2006}. From the data presented in figure 4 of that study, one can see that at physiological temperatures the characteristic times of base pair opening fluctuations are on the scale of tenths of ns for AT homopolymers and tens of ns for GC homopolymers, in accordance to the corresponding relaxation time scales obtained in the present work for the smaller widths. Further, looking at the shape of the local autocorrelation functions for T close to physiological temperature in~\cite{Kalosakas2006}, weak oscillations are present in the timescale 1-10 ps, while it is evident the much faster decay in the case of AT as compared to the GC sequences.

The substantial difference in bubble relaxation timescales between AT and GC homopolymers is a clear indication that the underlying base pairing dynamics play a strong role in determining the long-lasting effects of large out-of-equilibrium bubbles on the molecule.
At thermal equilibrium we have found that individual base pair opening fluctuations may on average live longer in the softer AT homopolymers as compared to the GC ones~\cite{Hillebrand2020}. However, that result concerned bubbles of larger amplitude in the AT sequences than in the GC sequences, while both of these amplitudes were one order of magnitude smaller than the heights considered here.

% section results (end)

%=========================
\section{Conclusions}
\label{sec:conclusions}
Within the framework of the Peyrard-Bishop-Dauxois model, we have computed the characteristic relaxation times for large bubbles in DNA homopolymers, using energy autocorrelation functions to study the equilibration of these coherent initial perturbations.
Through simulations of up to five microseconds, using efficient symplectic integration techniques, we ensure statistical accuracy of our results by averaging over many independent simulations (of the order of thousands).
By varying the initial bubble height and width in both pure AT and pure GC DNA sequences, and computing autocorrelation functions for each case, we found that the decaying relaxation dynamics, after some initial oscillations, consistently develops according to a stretched exponential evolution, matching the complex temporal behavior of these autocorrelation function (Fig.~\ref{fig:dataplots}).
The mechanism of the whole relaxation dynamics is related to a two stage process where the initial bubble is first rearranged towards an inherent localised structure, and then this more stable structure eventually spreads and completely decays to equilibrium (Fig.~\ref{fig:colourplots}).

The autocorrelation functions have distinct average relaxation times, calculated through the parameters of the stretched exponential fittings (Fig.~\ref{fig:params}), that depend on both bubble height and width, with larger amplitude and wider bubbles exhibiting longer relaxation times.
Computing the average relaxation times as a function of initial bubble height $h$ and width $w$ enables the direct quantification of the dependence of $\tau_{av}$ on $h$ and $w$ through linear fittings (Figs.~\ref{fig:tav}, \ref{fig:tavParams} and Eqs.~\ref{eq:tavGC}, \ref{eq:tavAT}).

The relaxation timescales for GC homopolymers are typically over two orders of magnitude longer than those evident in AT homopolymers;  within the used model, GC relaxation times range between $80-200$ ns, while the AT relaxations were on the order of $0.4-1$ ns, for bubble amplitudes up to 5.5~\AA\ and widths up to 21 base pairs that have been considered here.
These findings demonstrate that large bubbles leave significant imprints on the long-term dynamics of DNA molecules, and the extent of this impact depends strongly on the base pair composition of the sequence.

A subsequent continuation of this work would be to study the effect of heterogeneity in the DNA sequence on the characteristic relaxation times.
Further, it would be interesting to investigate how these dynamics develop in functional gene promoter regions.

%=========================
\section*{Acknowledgments} % (fold)

M.~H.~acknowledges support by the National Research Foundation (NRF) of South Africa. G.~K.~was supported by the Erasmus+/International Credit Mobility KA107 program. We thank the High Performance Computing facility of the University of Cape Town and the Center for High Performance Computing of South Africa for providing computational resources for this project.

%=========================
%_______________________________________________


\begin{thebibliography}{99}

\bibitem{beece1980} D.~Beece et al, Biochemistry \textbf{19}, 5147 (1980) .

\bibitem{ansari1985} A.~Ansari et al, Proc.~Natl.~Acad.~Sci.~USA \textbf{82}, 5000 (1985)

\bibitem{sobel1985} H.M. Sobell, Proc.~Natl.~Acad.~Sci. {\bf 82}, 5328 (1985).

\bibitem{Somoza2004} M.M.~Somoza, D.~Andreatta, C.J.~Murphy, R.S.~Coleman and M.A.~Berg, Nucleic Acids Res., \textbf{32}, 2494 (2004).

\bibitem{Perez2007} A.~P\'erez, F.~Javier Luque and M.~Orozco, J.~Am.~Chem.~Soc., \textbf{129}, 14739 (2007).

\bibitem{Banerjee2007} D.~Banerjee and S.K.~Pal, J.~Phys.~Chem.~B, \textbf{111}, 10833 (2007).

\bibitem{Galindo2014} R.~Galindo-Murillo, D.R.~Roe and T.E.~Cheatham III, Nat.~Comm., \textbf{5}, 5152 (2014).


\bibitem{Parmar2016} J.J.~Parmar, D.~Das, R.~Padinhateeri, Nucleic Acids Res., \textbf{44}, 1630 (2016).


\bibitem{Zeng2006} Y.~Zeng and G.~Zocchi, Biophys.~J., \textbf{90}, 4522 (2006).

\bibitem{leroy1988} J.L.~Leroy, M.~Kochoyan, T.~Huynh-Dinh and M.~Gu\'eron, J.~Mol.~Biol.,\textbf{200}, 223 (1988).

\bibitem{Jose2012} D.~Jose, S.E.~Weitzel and P.~H.~von Hippel, Proc.~Natl.~Acad.~Sci.~USA, \textbf{109}, 14428 (2012).

\bibitem{Phelps2013} C.~Phelps, W.~Lee, D.~Jose, P.H.~von Hippel, A.H.~Marcus, Proc.~Natl.~Acad.~Sci.~USA, \textbf{110}, 17320 (2013).

\bibitem{Lavery2010} Lavery et al, Nucleic Acids Res., \textbf{38}, 299 (2010).

\bibitem{Peyrard2009} M.~Peyrard, S.~Cuesta-L\'opez and G.~James, J.~Biol.~Phys. \textbf{35}, 73 (2009).

\bibitem{Manghi2016} M. Manghi and N. Destainville, Phys. Rep. {\bf 631}, 1 (2016).

\bibitem{kaxiras12} C.W. Hsu, M. Fyta, G. Lakatos, S. Melchionna, and E. Kaxiras, J.~Chem.~Phys.~{\bf 137}, 105102 (2012).

\bibitem{weber06} G. Weber, N. Haslam, N. Whiteford, A. Prugel-Bennett, J.W. Essex, and C. Neylon, Nature Phys. {\bf 2}, 55 (2006).

\bibitem{zoli13}  M. Zoli, J. Chem. Phys. {\bf 138}, 205103 (2013).

\bibitem{depablo14}  G.S. Freeman, D.M. Hinckley, J.P. Lequieu, J.K. Whitmer, and J.J. de Pablo, J. Chem. Phys. {\bf 141}, 165103 (2014).

\bibitem{zoli18}  M. Zoli, Physica A {\bf 492}, 903 (2018).

\bibitem{zoli18b}  M. Zoli, J. Chem. Phys. {\bf 148}, 214902 (2018).

\bibitem{sun19} X. Wang and Z. Sun, J. Chem. Inf. Model. {\bf 59}, 2980 (2019).

\bibitem{Peyrard1989} M. Peyrard and A.R. Bishop, Phys. Rev. Lett. {\bf 62}, 2755 (1989).

\bibitem{PBD} T. Dauxois, M. Peyrard, and A.R. Bishop, Phys. Rev. E {\bf 47}, R44 (1993).

\bibitem{DPB93} T. Dauxois, M. Peyrard, and A.R. Bishop, Phys. Rev. E {\bf 47}, 684 (1993).

\bibitem{Dauxois1995} T. Dauxois and M. Peyrard, Phys. Rev. E {\bf 51}, 4027 (1995).

\bibitem{Campa1998} A. Campa and A. Giansanti, Phys. Rev. E {\bf 58}, 3585 (1998).

\bibitem{Peyrard2000}  M. Peyrard and J. Farago, Physica A {\bf 288}, 199 (2000).

\bibitem{Kalosakas2005} G. Kalosakas, K.L. Ngai, and S. Flach, Phys. Rev. E {\bf 71}, 061901 (2005).

\bibitem{Kalosakas2011} G. Kalosakas, Phys. Rev. E {\bf 84}, 051905 (2011).

\bibitem{Chetverikov2019} A.P. Chetverikov, W. Ebeling, V.D. Lakhno and M.G. Velarde, Phys. Rev. E \textbf{100}, 052203 (2019).

\bibitem{Barre2001} J. Barre and T. Dauxois, Europhys. Lett. {\bf 55}, 164 (2001).

\bibitem{Hillebrand2019} M. Hillebrand, G. Kalosakas, A. Schwellnus, Ch. Skokos, Phys. Rev. E \textbf{99}, 022213 (2019).

\bibitem{Muniz2020} M.I. Muniz, H.H. Lackey, J.M. Heemstra and G. Weber, Chem. Phys. Lett. \textbf{749} 137413 (2020).

\bibitem{Behnia2020} S.~Behnia, S.~Fathizadeh, E.~Javanshour, and F.~Nemati, J.~Phys.~Chem.~B, \textbf{124}, 3261 (2020).

\bibitem{Alexandrov2009ePBD} B.S. Alexandrov, V. Gelev, Y. Monisova, L.B. Alexandrov, A.R. Bishop, K.\O. Rasmussen, and
A. Usheva, Nucleic Acids Res. {\bf 37}, 2405 (2009).

\bibitem{Voulgarakis2004} N.K. Voulgarakis, G. Kalosakas, K.\O. Rasmussen, and A.R. Bishop, Nano Lett. {\bf 4}, 629 (2004).

\bibitem{Ares2005} S. Ares, N.K. Voulgarakis, K.\O. Rasmussen, and A.R. Bishop, Phys. Rev. Lett. {\bf 94}, 035504 (2005).

\bibitem{Ares2007} S. Ares and G. Kalosakas, Nano Lett. {\bf 7}, 307 (2007).

\bibitem{Kalosakas2009} G. Kalosakas and S. Ares, J. Chem. Phys. {\bf 130}, 235104 (2009).

\bibitem{falo10}  R. Tapia-Rojo, J.J. Mazo, and F. Falo, Phys. Rev. E {\bf 82}, 031916 (2010).

\bibitem{Kalosakas2006} G. Kalosakas, K.\O. Rasmussen, and A.R. Bishop, Chem. Phys. Lett. {\bf 432}, 291 (2006).

\bibitem{Choi2004} C.H. Choi, G. Kalosakas, K.\O. Rasmussen, M. Hiromura, A. Bishop, and A. Usheva,
Nucleic Acids Res. {\bf 32}, 1584 (2004).

\bibitem{Kalosakas2004} G. Kalosakas, K.\O. Rasmussen, A.R. Bishop, C.H. Choi, and A. Usheva,
Europhys. Lett. {\bf 68}, 127 (2004).

\bibitem{Alexandrov2006} B.S. Alexandrov, L.T. Wille, K.\O. Rasmussen, A.R. Bishop and K.B. Blagoev, Phys. Rev. E \textbf{74},050901 (2006).

\bibitem{Choi2008} C.H. Choi, Z. Rapti, V. Gelev, M.R. Hacker, B. Alexandrov, E.J. Park, J.S. Park, N. Horikoshi,
A. Smerzi, K.\O. Rasmussen, A.R. Bishop, and A. Usheva, Biophys. J. {\bf 95}, 597 (2008).

\bibitem{Alexandrov2009} B.S. Alexandrov, V. Gelev, S.W. Yoo, A.R. Bishop, K.\O. Rasmussen, and A. Usheva,
PLoS Comput. Biol. {\bf 5}, e1000313 (2009).

\bibitem{Alexandrov2010} B.S. Alexandrov, V. Gelev, S.W. Yoo, L.B. Alexandrov, Y. Fukuyo, A.R. Bishop,
K.\O. Rasmussen, and A. Usheva, Nucleic Acids Res. {\bf 38}, 1790 (2010).

\bibitem{Apostolaki2011} A. Apostolaki and G. Kalosakas, Phys. Biol. {\bf 8}, 026006 (2011).

\bibitem{faloPRE12}  R. Tapia-Rojo, D. Prada-Gracia, J.J. Mazo, and F. Falo, Phys. Rev. E {\bf 86}, 021908 (2012).

\bibitem{huangJBE} H.-H. Huang and P. Lindblad, J. Biol. Eng. {\bf 7}, 10 (2013).

\bibitem{faloPLOS}  R. Tapia-Rojo, J.J. Mazo, J.A. Hernandez, M.L. Peleato, M.F. Fillat and F. Falo,
PLoS Comput. Biol. {\bf 10}, e1003835 (2014).

\bibitem{Hillebrand2021} M.~Hillebrand, G.~Kalosakas, A.R.~Bishop and Ch.~Skokos, J.~Chem.~Phys.~\textbf{155}, 095101 (2021).

\bibitem{Hillebrand2020} M.~Hillebrand, G.~Kalosakas, Ch.~Skokos and A.R.~Bishop, Phys.~Rev.~E \textbf{102} 062114 (2020).

\bibitem{Hairer2002} E. Hairer, C. Lubich, and G. Wanner, Geometric Numerical Integration, Vol. \textbf{31} (Springer, New York) 2002.

\bibitem{DMMS19} C.~Danieli, B.~Many Manda, T.~Mithun and  Ch.~Skokos, Math. in Engineering, \textbf{1},  447 (2019).

\bibitem{Blanes2002} S. Blanes and P. Moan, Journ. Comp. App. Math., \textbf{142}, 313 (2002).

\bibitem{Ngai1984} K.L.~Ngai, R.W.~Rendell and H.~Jain, Phys.~Rev.~B \textbf{30}, 2133 (1984).

\bibitem{Leon1997} C.~L\'eon, M.L.~Luca and J.~Santamaria, Phys.~Rev.~B \textbf{55}, 882 (1997).

\end{thebibliography}
\end{document}